\documentclass[onecolumn,epjc3]{svjour3} 

 \RequirePackage{mathptmx}     
\RequirePackage{latexsym}
\RequirePackage[numbers]{natbib}
\RequirePackage[colorlinks,citecolor=blue,urlcolor=blue,linkcolor=blue]{hyperref}

\journalname{Eur. Phys. J. C}

\usepackage[a4paper, total={6in, 9in}]{geometry}
\usepackage[utf8]{inputenc}

\usepackage{amssymb} 
\usepackage{amsmath} 
\usepackage{graphicx}

\usepackage{soul}  
\usepackage{ulem}
 
\newcommand{\mG}{\mathcal{G}}
\newcommand{\mL}{\mathcal{L}}
\newcommand{\mO}{\mathcal{O}}

\usepackage{color}

\begin{document} 

\title{Toy models for hierarchy studies}
\author{
    Clara \'Alvarez-Luna\thanksref{e1,addr1} 
    \and 
    Jos\'e A. R. Cembranos\thanksref{e2,addr1} 
    \and
    Juan Jos\'e Sanz-Cillero\thanksref{e3,addr1} 
    }

\thankstext{e1}{e-mail: c.a.luna@ucm.es}
\thankstext{e2}{e-mail: cembra@ucm.es}
\thankstext{e3}{e-mail: jjsanzcillero@ucm.es}

\institute{ Departamento de F\'isica Te\'orica and IPARCOS, Facultad de Ciencias F\'isicas, \\
Universidad Complutense de Madrid, 28040 Madrid, Spain \label{addr1}
}

\date{\today }

\maketitle

\begin{abstract}

We provide a simple computation in order to estimate the probability of a given hierarchy between two scales. 
In particular, we work in a model provided with a gauge 
symmetry, with two scalar doublets.  We start from a scale-invariant classical Lagrangian, but by taking 
into account the Coleman-Weinberg mechanism, we obtain masses for the gauge bosons and the scalars. 
This approach typically provides a {\it light} ($L$) and a {\it heavy} ($H$) sector related to the two 
different vacuum expectation values of the two scalars. 
We compute the size of the hypervolume of 
the parameter space of the model associated with an interval of mass ratios between these two sectors. 
We define the probability as proportional to this size and conclude that probabilities of very large 
hierarchies are not negligible in the type of models studied in this work.

\end{abstract}

\maketitle

\setcounter{equation}{0}

\section{Introduction}

There exist different fundamental energy scales within our present knowledge of physics. 
From a simplified point of view, we can 
refer to the reduced Planck scale $M_P = 10^{18}$ GeV as the one that suppresses the non-renormalizable gravitational 
interactions; the electro-weak scale $M_{EW}= 10^2$ GeV, as the one associated to the Higgs 
Vacuum Expectation Value (vev); the neutrino scale $M_{\nu}= 10^{-10}$ GeV, whose square is of the order
of the square difference of neutrino masses deduced from oscillation experiments; and the cosmological 
constant scale $M_{\Lambda}= 10^{-12}$~GeV, that in the standard cosmological model is the scale related to 
the negative pressure necessary to accelerate the late expansion of the Universe.

One of the main fundamental questions in theoretical physics is to understand the origin of such scales.
Indeed, we are not sure if these scales can be explained in terms of more fundamental physics, or they are
given as {\it inexplicable} fundamental parameters. 

From the point of view of the Quantum Field Theory (QFT), the situation is more involved. Typically, the above 
scales are determined by fundamental constants in the action of the theory. For instance, 
$M_P = 10^{18}$~GeV is related to the Newton constant that appears in front of the Einstein-Hilbert action
in General Relativity (GR); $M_{EW}$ is given by the constant associated to the quadratic term of the Higgs doublet in the Standard Model (SM); $M_{\Lambda}$ is fixed by the cosmological constant; and
$M_{\nu}$ is associated with the mass term of the active neutrinos
\footnote{Although this scale, as the others, could be the result
of other fundamental scales. This is the case in the well-known {\it See-Saw} (SS) models. Roughly speaking, 
the SS scale is given by the mass of one or several heavy sterile neutrinos, in such a way that 
$M_{SS}\simeq M_{EW}^2/M_{\nu}\simeq 10^{14}$ GeV.}.

In QFT, the above constants suffer from the renormalization prescription. It means that the observed values measured in experiments are not just given by the bare constants that appear in the action, but they contain radiative corrections~\cite{Coleman:1973jx,Casas:1994qy,Buttazzo:2013uya}. 
Indeed, if one is not careful enough in the construction or extension of a particular QFT, 
these corrections can be so important that can lead to the so-called quantum instabilities and fine-tuning problems~\cite{Ellis:1986yg,Barbieri:1987fn,Ciafaloni:1996zh,Casas:2014eca}.    
This is the most important theoretical problem associated with the large hierarchies between some of the above commented scales \cite{Weinberg:1988cp,Bardeen:1995kv,Fichet:2012sn}.

From a different approach, the energy scales may not be present in the action, but they can be originated by the quantum corrections themselves. Although the classical theory were scale-invariant, this symmetry would be anomalous. The possibility of producing all the fundamental scales by quantum effects have been pursued in different frameworks, with the Coleman-Weinberg effective potential mechanism one of the most popular approaches~\cite{Coleman:1973jx}.  In this work, we do not try to give a viable solution to this question, but rather study the general problem of the existence of large scale hierarchies. We will illustrate the issue with a toy model and estimate the probability of generating a large separation of scales from quantum origin. Alternative probabilistic analyses of the naturalness problem can be found in Refs.~\cite{Cabrera:2008tj,Fichet:2012sn,Ghilencea:2012qk}. However, the latter are typically based in Bayesian studies that lead to a Barbieri-Giudice measure-type \cite{Barbieri:1987fn}. These approaches determine the probability of a model under the assumption of some set of data and prior distributions. In contrast, we focus on the calculation of the likelihood of having a given mass ratio for concrete values of the model parameters. We find a logarithmic suppression at large hierarchies that is not present in previous studies. This feature is related to the absence of dimensionful parameters in the particular model under study. In this sense, this type of theories seems to be promising for alleviating naturalness problems.

Our proposal is to start from a massless Lagrangian and make use of the Coleman-Weinberg (CW) 
mechanism~\cite{Coleman:1973jx,Weinberg:1973am,Gildener:1976ih,Chataignier:2018kay,Khoze:2016zfi}   
to obtain Spontaneous Symmetry Breaking (SSB) and generate the mass scales of the model. These authors~\cite{Coleman:1973jx} showed how a theory that is symmetric when looking at the interactions present in the tree level Lagrangian can develop SSB when the radiative corrections are taken into account. Thus it is possible to generate masses for some particles, even in theories that do not explicitly include any energy scale~\footnote{By contrast, the Standard Model Higgs potential contains an explicit mass scale and the electroweak SSB is triggered at tree-level, with quantum loops introducing small subdominant corrections.}. The basic CW approach~\cite{Coleman:1973jx} considers only one-loop corrections, but higher order contributions can also be studied~\cite{Buttazzo:2013uya}. Within the CW mechanism, quantum loops yield the dominant contributions that generate the SSB. 

\section{$SU(2)_L \times SU(2)_H \times U(1)_X$  model}

We will work with one of the simplest models that can provide two different scales from radiative corrections: we will assume a model with a gauge symmetry group $\mG=SU(2)_L \times SU(2)_H \times U(1)_X$ containing two complex scalar doublets under $SU(2)_L$ and $SU(2)_H$, $\Phi$ and $\Theta$, respectively. Thus, we will have the following particle content: the $SU(2)_L$ gauge boson triplet  $(W_L^{1\mu}, W_L^{2\mu}, W_L^{3\mu})$; the  $SU(2)_H$ gauge boson triplet $(W_H^{1\mu}, W_H^{2\mu}, W_H^{3\mu})$; the $U(1)_{X}$ gauge boson singlet $X^\mu$; one (light sector) $SU(2)_L$ scalar complex doublet $\Phi$; and one (heavy sector) $SU(2)_H$ scalar complex doublet $\Theta$. These doublets $\Phi$ and $\Theta$ have, respectively, the Abelian charges $Q_L$ and $Q_H$ under $U(1)_X$, being both sectors connected by the $X^\mu$ gauge boson. Eventually, without any loss of generality, it will be useful to choose the orientation of these scalar fields as $\Phi^T=(0,\varphi)/\sqrt{2}$ and $\Theta^T=(0,\eta)/\sqrt{2}$.  

The interactions in this model will be provided by the renormalizable Lagrangian,  
\begin{equation}
\mL_0 \,=\, |D_\mu\Phi|^2 +|D_\mu\Theta|^2 - V_0\, , 
\end{equation}
which includes the potential, 
\begin{equation}
V_0(\varphi, \eta) = \frac{1}{4!} \lambda_L\varphi^4 
+\frac{1}{4!} \lambda_H\eta^4 
+\frac{1}{4!} \lambda_{LH}\varphi^2\eta^2 \,, 
\end{equation}
with $\varphi^2=2|\Phi|^2$, $\eta^2=2|\Theta|^2$.  

The covariant kinetic term provides the gauge boson mass terms. In the $LH$ decoupled limit with $g_X=0$, one has $m_{W_{L,\, j}}=g_L \varphi/2 $ and  $m_{W_{H,\, j}}=g_H \eta/2 $ (with $j=1,2,3$) and $m_{X}=0$.  
For $g_X\neq 0$ the expressions of the masses are a bit more involved: 
$W^\mu_{L,\, 1} $, $W^\mu_{L,\, 2} $, $W^\mu_{H,\, 1} $ and $W^\mu_{H,\, 2}$ masses are the same as in the $g_X=0$ case but a mixing shows up between the $W^\mu_{L,\, 3}$, $W^\mu_{H,\, 3}$ and $X^\mu$ gauge bosons, leading to the diagonalized mass eigenstates $Z^\mu_L,\, Z^\mu_H$ and $\hat{\gamma}^\mu$. An eigenstate, $\hat{\gamma}^\mu$, is always massless while the $Z_L^\mu$ and $Z^\mu_H$ masses depend on a combination of the three gauge couplings.~\footnote{ 
The masses of the $Z^\mu_L$ and $Z^\mu_H$ gauge bosons are given by, 
\begin{eqnarray}
&&m_{Z_{L,H}}^2= \frac{\overline{M}^2}{2} \left[ 1 \mp \sqrt{1- 4\overline{m}^2/\overline{M}^2} \right] \, ,
\quad 
\overline{M}^2=
(g_H^2 + g_X^2 Q_H^2 )\eta^2 + (g_L^2 + g_X^2 Q_L^2) \varphi^2\, ,
\quad 
\overline{m}^2=   
  ( g_ H^2 g_ L^2 + g_X^2 (  Q_H^2 g_ L^2 + Q_L^2 g_ H^2)) \eta^2 \varphi^2/(4 \overline{M}^2)\, ,  
\nonumber
\end{eqnarray}
with $m_{Z_L}\simeq \overline{m}$ and $m_{Z_H}\simeq \overline{M}$ for large hierarchies ($\overline{m}\ll \overline{M}$).
}  
In any case, in the $g_X\to 0$ limit, one has $Z^\mu_{L}\to W^\mu_{L,\, 3}$, $Z^\mu_{H}\to W^\mu_{H,\, 3}$  and $\hat{\gamma}^\mu\to X^\mu$, as expected.    

Finally we can construct the effective potential, including the tree level terms and logarithmic one-loop corrections. In the Coleman-Weinberg approximation, scalar loops are assumed to be negligible with respect to the gauge boson ones. 
Thus, the one-loop corrections are determined by the gauge bosons masses in the form~\cite{Coleman:1973jx},  
\begin{equation}
V = V_0 + \frac{3}{64\pi^2} \sum_{j=1}^7 m_j^4\left[ \ln\left(\frac{m_j^2}{\mu^2}\right)-\frac{5}{6}\right]\,, 
\label{eq:CW-potential}
\end{equation}
with $m_j(\varphi,\eta)$ the masses of each of the seven $SU(2)_L\times SU(2)_H\times U(1)_X$ gauge bosons  and with $\mu$  the renomalization scale in the $\overline{MS}$ scheme. The $m_j(\varphi,\eta)$ functions depend on the value of the scalar fields $\varphi$ and $\eta$ and become the physical gauge boson masses at the vev of these scalar fields, $\langle \varphi\rangle$ and $\langle \eta\rangle$, respectively. 
We want to study different cases and limits, depending on the values of the different couplings. Within these cases, we will analyze 
the probability of obtaining a given mass hierarchy. For that we define the following ratio:
\begin{equation}
    \mathfrak{R} \,\, =\,\,  \frac{g_H^2 \langle\eta\rangle^2}{g_L^2\langle\varphi\rangle^2}
    \,\,=\,\, 
    \frac{m_{W_H}^2 (\langle\eta\rangle) }{m_{W_L}^2 (\langle\varphi\rangle) }
    \ ,
\end{equation}
which gives us the hierarchy between the square masses of the (non-mixed) heavy gauge bosons ($H$) and light gauge bosons ($L$). 
We are mainly interested in large hierarchies in the masses without large hierarchies between the different dimensionless 
couplings: in this scenario a large hierarchy between $m_{W_H}$ and $m_{W_L}$ is equivalent to a large hierarchy between $m_\eta$ and $m_\varphi$. This ratio $\mathfrak{R}$ is simply more convenient for the analytical derivation below. 
Therefore, the above ratio is a good parameter not only to estimate the hierarchy between
the masses of the two gauge bosons, but also between the complete two sectors. 

The phenomenology of the model depends on the values of its different couplings. To make the analysis simpler, we will study various scenarios, with different parameters set to zero. This will allow us to separate the contributions of each coupling to the potential and study its implications to the hierarchies of the model. The CW potential discussion in Ref.~\cite{Chataignier:2018kay} for a closely related $SU(2)\times SU(2)$ model can be useful for further clarifications, as it explores alternative situations.  

We are considering two main restrictions that limit the possible values of the parameters in order to have a consistent model:  
\begin{eqnarray}
\mbox{CW approx.: } &\qquad& |\lambda_j| \, < \,  \epsilon_{CW}\, \cdot \,  g_j^2 \, , 
\nonumber\\
\mbox{perturbative: } &\qquad& g_j^2 \, < \,  \epsilon_{g^2} \,\cdot\,  4\pi\,  \equiv\,  g^2_{max}\ ,
\label{eq:restrictions}
\end{eqnarray}
where the different $\epsilon_i\ll 1$ are the tolerances associated to each restriction. 
The first constrain ensures that $\lambda_j^2\ll g_j^4$ and, therefore, the validity of the CW approximation~\cite{Coleman:1973jx}, where radiative corrections are fully dominated by gauge boson loops (scalar loops are neglected). 
The second restriction implies that 
${g_j^2}/{(4\pi)}\ll 1$, so higher loop corrections can be safely ignored.  
Regarding perturbativity, in principle, one might also consider a third constraint $|\lambda_j| < \epsilon_\lambda \, \cdot \, 4\pi$,  
in such a way that $\epsilon_\lambda\ll 1$ ensures 
${\left|\lambda_j \right|}/{(4\pi)}\ll 1$.
However, the fulfilment of the first two conditions immediately implies perturbativity in the $\lambda_j$ expansion and, hence, it will be no longer discussed. 
These restrictions define a triangular region in our $(\lambda_H,g_H^2)$ parameter space --and similarly in the $(\lambda_L,g_L^2)$ plane--, in which we will study the different hierarchies. 
The precise value of the tolerance parameters has no large effects on the results as far as both are of a similar order  
(in fact, they do not play any role if $\epsilon_{g^2}=\epsilon_{CW}$).

\begin{figure}[!t]
\begin{center}
\includegraphics[scale=0.575]{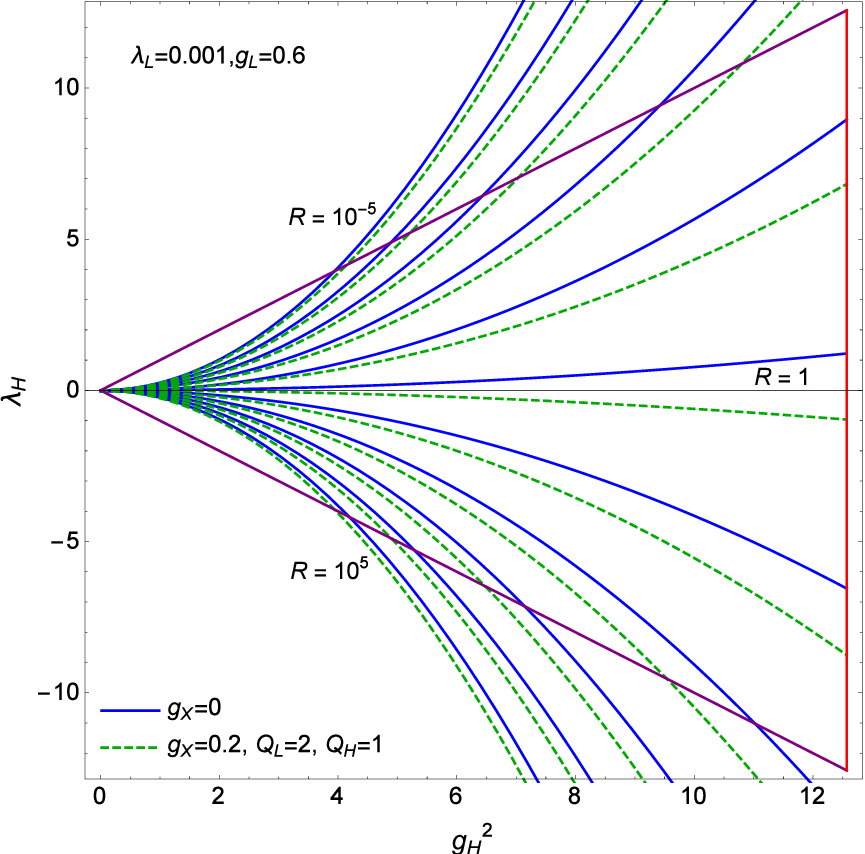}
\caption{{\small Illustration of the allowed parameter region in the $(g_H^2,\lambda_H)$ plane  
and the lines with constant $\mathfrak{R}$ for a given value of $g_L$ and $\lambda_L$.
For illustration, we show large hierarchies up to $\mathfrak{R}=10^5$ (lower region of the plot) and small hierarchies up to $\mathfrak{R}=10^{-5}$ (upper region of the plot), while the hierarchy $\mathfrak{R}=1$ remains in the middle of the plot. 
From top to bottom, each line increases its $\mathfrak{R}$ value by a factor $\times 10$.
The solid blue lines correspond to the results with $g_X=0$ while the dashed green ones correspond to $g_X=0.2$, $Q_L=2$ and $Q_H=1$. The restrictions described in Eq.~(\ref{eq:restrictions})  
are also represented
for $\epsilon_{CW}=\epsilon_{g^2}=1$}: CW restriction (diagonal purple lines) and $g^2_{max}$ (vertical red line).   }
\label{fig:area}
\end{center}
\end{figure}

\section{Hierarchy probabilistic analysis}
 
Our first approach consists on fixing the couplings of the $L$ sector instead of integrating the whole $L$--$H$ space of parameters. 
We study the conditional probability for a set of given $L$--couplings $\alpha_L=\{g_L^2, \lambda_L\}$. One can observe this as a scenario where we have a certain fixed knowledge of the theory at low energies but still consider all possibly allowed configurations for the $H$ sector couplings. 
In this case, this conditional probability $\mathfrak{P}^{(\alpha_L)}$ of being between the hierarchies $\mathfrak{R}_0$ and $\mathfrak{R}_1$ is proportional to the area between the curves with constant $\mathfrak{R}_0$ and $\mathfrak{R}_1$ contained in the allowed $(g_H^2,\lambda_H)$ region. 
For the decoupled scenario ($g_X=0$, $\lambda_{LH}=0$), the lines of constant $\mathfrak{R}$ in the $(g_H^2,\lambda_H)$ plane are given by 
\begin{equation}
\label{eq:hierarchy1}
    \mathfrak{R} = e^{\frac{128\pi^2}{27}\left(\frac{\lambda_L}{g_L^4}-\frac{\lambda_H}{g_H^4}\right)}\ .
\end{equation}
If we allow $g_X\neq 0$ (though keeping $\lambda_{LH}=0$ at the given $\mu$), the curve of constant $\mathfrak{R}$ still remains simple enough to be dealt with analytically for the small $g_X$ expansion of $V(\varphi,\eta)$ in~(\ref{eq:CW-potential}):
\begin{equation}
\mathfrak{R} = e^{\frac{128\pi^2}{27}\left( \frac{\left(\lambda_L\, +\, \frac{9}{128\pi^2} g_L^2Q_L^2 g_X^2\right)}{\left(g_L^4\, +\, \frac{2}{3} g_L^2Q_L^2 g_X^2\right)} \,-\, (L\leftrightarrow H) \right) }\ .
\label{eq:R-for-gXnot0}
\end{equation}

The case $\lambda_{LH}\neq 0$ is a little more involved. It admits an implicit analytical relation between $\langle\varphi\rangle$ and $\langle\eta\rangle$ which is provided in~\ref{app:lambdaLH-corrections} for $g_X=0$. 
A non-zero value of $\lambda_{LH}$ modifies the relations that determine the vev's and the $\mathfrak{R}$ hierarchy ratio in  Eq.~(\ref{eq:hierarchy1}). For $\mathfrak{R}>1$, a positive $\lambda_{LH}$ always tends to make $\mathfrak{R}$ even larger while, for negative $\lambda_{LH}$, the effect is just the opposite, reducing the value of $\mathfrak{R}$. For the case with $\mathfrak{R}<1$, these two behaviours are interchanged: a positive $\lambda_{LH}$ makes $\mathfrak{R}$ smaller, whereas a negative $\lambda_{LH}$ makes it larger. Therefore, the measure of the parameter-space hypervolume with large hierarchies will not be drastically modified: the qualitative results later derived will not change. However, the mathematical discussion is much more involved and less straightforward. For this reason, for the study of the coupling between the $L$ and $H$ sectors, we will focus on $g_X\neq 0$ perturbations but with $\lambda_{LH}= 0$.

One can see that, even if we introduce small differences between the couplings in the light ($\lambda_L,g_L$) and the heavy sectors ($\lambda_H,g_H$), 
both huge or very small hierarchies can be generated between them due to the exponential factor 
(notice $e^{128\pi^2/27} \sim  10^{20}$). Likewise, we note that if there is a given hierarchy $\mathfrak{R}$ (this is, between the gauge bosons masses), the same approximate hierarchy appears between the vev's and between the physical scalar masses. On the other hand, if both gauge groups have exactly the same couplings with $g_L = g_H$ and $\lambda_L = \lambda_H$ in decoupled models, both vev's will be equal and the same will happen between the scalar and gauge boson masses of the $L$ and $H$ sectors, i.e., we would have $\mathfrak{R} = 1$. This is also generally true for coupled models in which $Q_L=Q_H$ with $g_X,\lambda_{LH}\neq 0$. Fig.~\ref{fig:area} shows the constant $\mathfrak{R}$ lines in the $(g_H^2,\lambda_H)$ allowed region for  $g_L=0.6$, $\lambda_L=10^{-3}$, $g_X=0$ and $\lambda_{LH}=0$, and also the shift that these lines suffer when 
mixing couplings $Q_L g_X, Q_H g_X\neq 0$ are included. The area of these regions can be integrated analytically without much problem. The conditional probability $\mathfrak{P}^{(\alpha_L)}$ is provided by the ratio of the area with $\mathfrak{R}\in[\mathfrak{R}_0,\mathfrak{R}_1]$ 
and the total allowed area in the $(g_H^2,\lambda_H)$ plane (given by the {\it CW-triangle} in Fig.~\ref{fig:area}). The cumulative probability from $\mathfrak{R}_0$ up to $\infty$ is given for the decoupled case ($g_X=\lambda_{LH}=0$) by~\footnote{
A global factor $(\epsilon_{CW}/\epsilon_{g^2})^2$ must be added to the results in~(\ref{eq:cond-prob-cumul_gX=0}) if different tolerances are considered ($\epsilon_{CW}\neq\epsilon_{g^2}$).
}
\begin{eqnarray}
&&\mathfrak{P}^{(\alpha_L)}_{\rm cumul}  
\, =\,  \frac{1}{6}\left(\frac{27\ln\mathfrak{R}_0}{32\pi} -\frac{4\pi\lambda_L}{g_L^4}\right)^{-2}
    \,\stackrel{\mathfrak{R}_0\gg 1}{\simeq} \, 
    \frac{0.44}{\left(\log_{10}\mathfrak{R}_0\right)^2}  \, ,
\nonumber
\\
\label{eq:cond-prob-cumul_gX=0}
\end{eqnarray}
for $\ln\mathfrak{R}_0> \frac{128\pi^2}{27}\left(\frac{\lambda_L}{g_L^4}+ \frac{\epsilon_{CW}}{4\pi\epsilon_{g^2}}\right)$. 
The form of $\mathfrak{P}^{(\alpha_L)}_{\rm cumul} $ for smaller $\mathfrak{R}_0$ can also be easily derived. However, this is not the $\mathfrak{R}$--range of interest in this article, so it will not be discussed in further detail. 
The differential probability to have a hierarchy within an interval  $\ln\mathfrak{R}\in[\ln\mathfrak{R}_0,\ln\mathfrak{R}_0+d\ln\mathfrak{R}_0]$ is in general related to the cumulative probability through 
\begin{eqnarray}
&&d\mathfrak{P}^{(\alpha_L)} 
\, = \, -\, \frac{  d\mathfrak{P}^{(\alpha_L)}_{\rm cumul}  }{d\ln\mathfrak{R}_0}  d\ln\mathfrak{R} 
\,\stackrel{\mathfrak{R}_0\gg1}{\simeq} \, 
    \frac{0.38  }{\left(\log_{10}\mathfrak{R}_0\right)^{3}} d\ln\mathfrak{R}\,,\nonumber\\
\label{eq:cond-prob-dens}
\end{eqnarray}
for the decoupled system. 

We have also studied the weakly coupled case with $g_X\neq 0$ (though keeping $\lambda_{LH}=0$, for simplicity). 
If the full $g_X$ contribution to the gauge boson masses $m_j^2\left(\langle\varphi\rangle,\langle\eta\rangle \right)$ is kept, the vev's $\langle\varphi\rangle$ and $\langle\eta\rangle$, which determine  $\mathfrak{R}$,   
can no longer be analytically computed and they have to be calculated numerically. Nevertheless, if one considers the perturbative expansion $m_j^2\approx   m_{j\, (0)}^2 + m_{j\, (2)}^2~g_X^2$ up to $\mO(g_X^2)$, we are able to extract the vev's and the analytical relation  $\mathfrak{R}=\mathfrak{R}(g_L,\lambda_L,g_H,\lambda_H,Q_L g_X, Q_H g_X)$ in Eq.~(\ref{eq:R-for-gXnot0}).

\begin{figure}[!t]
\begin{center}
\includegraphics[scale=0.575]{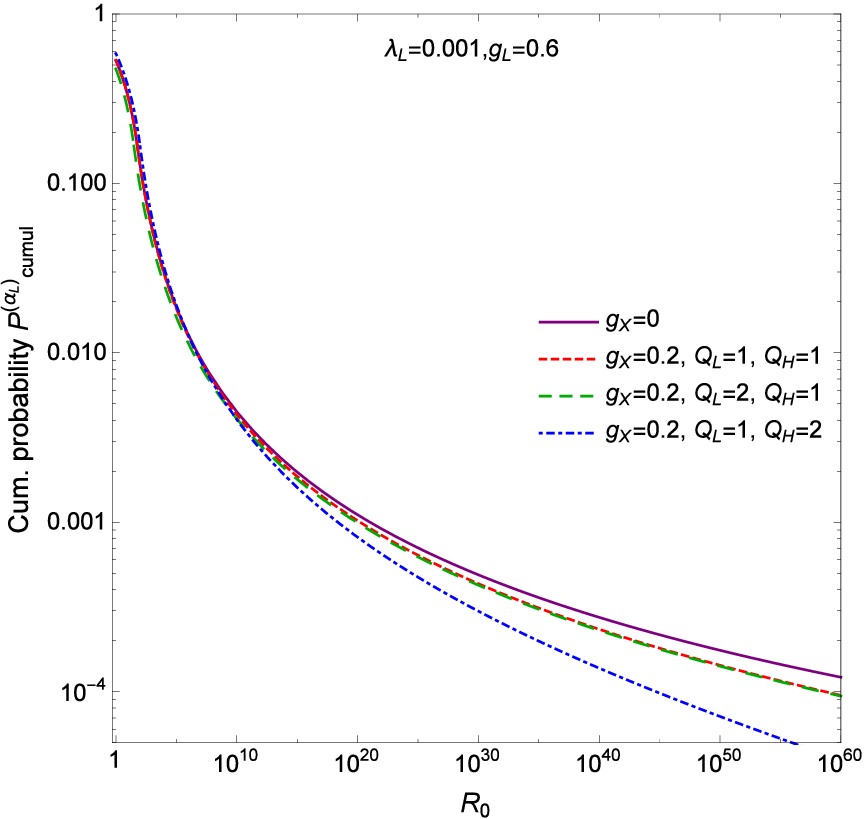}
\caption{{\small Comparison of the cumulative probability
$\mathfrak{P}_{\rm cumul}^{(\alpha_L)}$
for fixed $\lambda_L=10^{-3}$ and  ${g_L=0.6}$, both for decoupled $LH$ sectors ($g_X=0$) and weakly interacting $LH$ sectors (with $g_X=0.2$ and for different couplings $(Q_L,Q_H)$ for each sector).}}
\label{fig:prob-cumulVARIOS}
\end{center}
\end{figure}

In the case in which we fix the $\alpha_L=\{g_L^2,\lambda_L\}$ and $\{Q_L g_X, Q_H g_X\}$ parameters, it is possible to analytically compute the previously discussed area integral in the $(g_H^2,\lambda_H)$ plane. For sake of clarity, 
the expression for $\mathfrak{P}^{(\alpha_L)}$ is relegated to Eq.~(\ref{eq:cond-prob-cumul_gXnot0}) in~\ref{app:gX-corrections}. It is not difficult to observe that this result turns into~(\ref{eq:cond-prob-cumul_gX=0}) in the small $g_X$ limit. 
We illustrate these results in Fig.~\ref{fig:prob-cumulVARIOS}, where we provide the cumulative probability $\mathfrak{P}^{(\alpha_L)}_{\rm cumul}$ for $g_X=0$ (decoupled $LH$ sectors) and $g_X=0.2$ (weakly interacting $LH$ sectors, for different choices of $Q_L, Q_H$, with $\lambda_{LH}=0$). In all cases, we consider the inputs $\lambda_L=10^{-3}$ and  ${g_L=0.6}$. We note that the two curves with $g_X=0.2$ and $Q_H=1$ (red dashed, $Q_L=1$, and green long-dashed, $Q_L=2$) are very similar at large $\mathfrak{R}_0$ --though not identical-- (further details are given in~\ref{app:gX-corrections}). We find that the probability corrections with respect to the $g_X= 0$ case are tiny at small and intermediate hierarchies ($\mathfrak{R}\lesssim  10^{30}$). Nonetheless, although $\mathfrak{P}_{\rm cumul}^{(\alpha_L)}$ is not quite affected by the $g_X$ corrections for small $\mathfrak{R}$, things are different for very large $\mathfrak{R}$: we found that, asymptotically, the numerically computed cumulative probability with the full potential~(\ref{eq:CW-potential})  
($\mathfrak{R}\gtrsim  10^{30}$) 
disagrees with the prediction provided by the analytical perturbative $g_X$ expressions in Eq.~(\ref{eq:cond-prob-cumul_gXnot0}). 
Therefore, the shape of the $\mathfrak{R}\to\infty$ probability distribution becomes sensitive to the precise details of the ``weak'' interaction between the $L$ and $H$ sectors and one should go beyond perturbation theory in $g_X$.~\footnote{
We want to emphasize that, in order to test the validity of our approximations, all the analytical results in this article have been checked against numerical evaluations of the vev's $\langle\varphi\rangle$ and $\langle\eta\rangle$, and probabilities.  
We found that asymptotically,  for very large $\mathfrak{R}$, 
the vev's computed numerically from the full potential~(\ref{eq:CW-potential}) disagree with those obtained from the perturbative $g_X$ analytical expressions.
} 

\begin{figure}[!t]
\begin{center}
\includegraphics[scale=0.575]{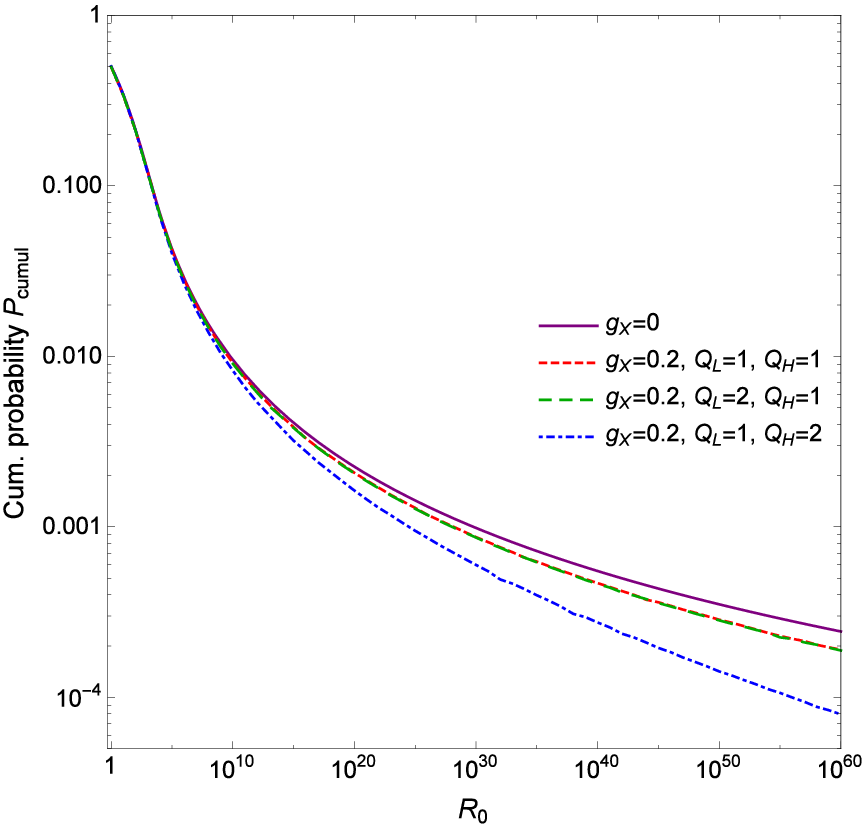}
\caption{{\small Comparison of the cumulative probability~(\ref{eq:prob-cumul}) for the integration of the whole $L$--$H$ hypervolume in the decoupled and coupled scenarios (with $g_X=0.2$ and for different couplings $(Q_L,Q_H)$ for each sector), both with $\lambda_{LH}=0$. 
}}
\label{fig:full-prob-cumul}
\end{center}
\end{figure}

The previous study of the $\alpha_L$--conditional probability $\mathfrak{P}_{\rm cumul}^{(\alpha_L)} $ shows that the decoupled scenario ($g_X=\lambda_{LH}=0$) seems to provide a fair enough approximation of the $L$--$H$ weakly interacting case for moderate hierarchies ($\mathfrak{R}\lesssim  10^{30}$). 
Hence, in what follows, our fully analytical approach will just focus on the decoupled limit for our last analysis, where we derive the $\mathfrak{R}$ probability distribution from the integration to the whole $(g_L^2,\lambda_L,g_H^2,\lambda_H)$ allowed parameter space.  
In addition, the coupled system with $g_X\neq 0$ is computed numerically by including corrections up to $\mathcal{O}(g_X^2)$ for the potential in Eq.~\eqref{eq:CW-potential}. 
The total probability distribution for weakly coupled models is plotted in Fig. \ref{fig:full-prob-cumul} and discussed below.

In the decoupled case ($g_X=\lambda_{LH}=0$), the hypervolume between two hypersurfaces with constant $\mathfrak{R}_0$  and $\mathfrak{R}_1$  can be analytically computed. Thus, it is possible to compare our numerical estimates with the exact analytical expression for the hierarchy  probability defined as the ratio between this hypervolume in the $L$--$H$ parameter space and the total allowed hypervolume. Notice that both the integrated volume and the total volume are proportional to $\epsilon^6$ in the case with identical tolerances $\epsilon_{CW}=\epsilon_{g^2}\equiv\epsilon$, so the probability calculated as their ratio turns out to be tolerance independent, as commented. The cumulative probability from $\mathfrak{R}_0$ up to $\infty$ is given by
~\footnote{
If the perturbativity and CW restrictions are taken to be different ($\epsilon_{CW}\neq \epsilon_{g^2}$) the results in Eqs.~(\ref{eq:prob-cumul}) and~(\ref{eq:prob-dens}) must be multiplied by a global factor $(\epsilon_{CW}/\epsilon_{g^2})^3$.   }
\begin{eqnarray}
&&    \mathfrak{P}_{\rm cumul} 
= \frac{
    1   
    }{3}\left(\frac{32\pi}{27\ln\mathfrak{R}_0}\right)^2  
\left[ 1 +   \left(\frac{32\pi}{27\ln\mathfrak{R}_0}\right)^2  
\right.
\left.
    \ln\left(\left(\frac{27\ln\mathfrak{R}_0}{32\pi}\right)^2-1\right)\right]
 \stackrel{\mathfrak{R}_0\gg 1}{\simeq}
    \frac{0.87 }{(\log_{10}\mathfrak{R}_0)^2}\, ,\;\;\;\;\;\;
\label{eq:prob-cumul}
\end{eqnarray}
for $\ln\mathfrak{R}_0> \frac{64\pi}{27}$.  
The form of $\mathfrak{P}_{\rm cumul} $ for smaller $\mathfrak{R}_0$ can also be computed without much problem.
Fig.~\ref{fig:full-prob-cumul} compares this result and those for $g_X\neq 0$, computed numerically. While all curves coincide for $\mathfrak{R}_0\sim 1$, small deviations appear as the hierarchy increases. 

At very large $\mathfrak{R}_0$, the small $g_X$ expansion of the potential fails and one must perform the analysis with the full CW potential~(\ref{eq:CW-potential}), not just the $g_X$ expansion up to $\mathcal{O}(g_X^2)$ shown in Fig.~\ref{fig:full-prob-cumul}.
Thus, large hierarchies are sensitive to the integration in the small $g_{L,H}$ and $\lambda_{L,H}$ range, where the $g_X$ parameter may become even dominant (in our fixed $g_X$ approach).   
Nonetheless, for not that large hierarchies ($\mathfrak{R}   \lesssim 10^{30}$) perturbation theory on $g_X$ works fine. Thus, when this perturbative expansion converges well, the constant $\mathfrak{R}$ lines in the $(g_H^2,\lambda_H)$ plane shown in Fig.~\ref{fig:area} are found to be close to those for $g_X=0$. Hence, we consider one can trust our results for the cumulative probability in that $\mathfrak{R}$ range. We note that, as it occurred with $\mathfrak{P}_{\rm cumul}^{(\alpha_L)}$ in Fig.~\ref{fig:prob-cumulVARIOS}, the two curves in Fig.~\ref{fig:full-prob-cumul} with $g_X=0.2$ and $Q_H=1$ (red dashed, $Q_L=1$, and green long-dashed, $Q_L=2$) are very similar at large $\mathfrak{R}_0$, although not exactly equal. 
The differential probability to have a hierarchy within an interval  $\ln\mathfrak{R}\in[\ln\mathfrak{R}_0,\ln\mathfrak{R}_0+d\ln\mathfrak{R}_0]$ is given for $g_X=0$ by,  
\begin{eqnarray}
    d\mathfrak{P}  
\, = \, -\, \frac{  d\mathfrak{P}_{\rm cumul}  }{d\ln\mathfrak{R}_0}
d\ln\mathfrak{R}  
    \,\stackrel{\mathfrak{R}_0\gg 1}{\simeq} \, 
    \frac{0.76  }{(\log_{10}\mathfrak{R}_0)^3} 
    d\ln\mathfrak{R}\,.    \;\;\;\;\;
\label{eq:prob-dens}
\end{eqnarray}

\section{Conclusions and final remarks}

In this article, we have studied a simple two doublet gauge model that allows large hierarchies between scales. We have constructed a probability that estimates how likely is to have a given hierarchy, defined by the hypervolume of the region of the parameter-space with that hierarchy $\mathfrak{R}_0$. We have considered two different situations for our toy model. First, we have estimated the probability to obtain a large hierarchy having a fixed low energy sector ($L$), analysing the impact of a weak coupling between both $L\leftrightarrow H$ sectors. Second, we have calculated the global probability to obtain a given hierarchy, scanning all possible values of both $L$ and $H$ sector parameters. Corrections due to a weak $L\leftrightarrow H$ mixing coupling do not change our basic conclusions. 

We conclude that a small hierarchy between sectors is more likely, since these hierarchies cover the largest regions of the parameter space; nevertheless, we have also shown that the decreasing of the probability for larger hierarchies is not sizeable, since the cumulative probability is only logarithmically suppressed, with  $\mathfrak{P}_{\rm cumul}\sim (\log_{10}\mathfrak{R})^{-2}$.
This behaviour is easy to understand observing Eq.~(\ref{eq:hierarchy1}), where there is an exponential dependence of the hierarchy on the couplings. As one can see in Fig.~\ref{fig:area}, when $\mathfrak{R}$ increases the area trapped below the constant-$\mathfrak{R}$ curve decreases, but at a logarithmic rate. This feature is expected in a wide set of theories with dynamical symmetry breaking, where the hypervolumes that determine the probabilities vary logarithmically with the hierarchy.   

Therefore, very large hierarchies are less probable but not as unlikely as one might {\it a priori} think. As an example, we can see that hierarchies of the order $\mathfrak{R}\gtrsim (M_P/M_{EW})^2\sim 10^{32}$ or even $\mathfrak{R}\gtrsim (M_P/M_{\Lambda})^2\sim 10^{60}$ would be only suppressed by probabilities $\mathfrak{P}_{\rm cumul}\sim 10^{-3}$--$10^{-4}$. In addition, we have also shown that these results are robust against ambiguities in the probability definition (tolerances) and the coupling parameters (at least, as far as they are not in the strongly coupled regime).

Finally, we would like to discuss how these results are expected to change if different models are considered. In particular, we can analyse different symmetry groups $\mG$. The  result in this article is indeed very general for models with symmetries with a more general product structure $\mG={ \prod_{\chi}} SU(2)_\chi$. In the limit when the various $\chi$ sectors are decoupled or weakly interacting, one can consider exactly the same arguments applied here for the $L$ and $H$ sectors. Thus, for any two sectors  $L,H\in \{\chi\}$ one would obtain for the hierarchy ratios $\mathfrak{R}\equiv m_{W,L}^2 / m_{W,H}^2$ similar probability distributions to those obtained in this work for the simpler case with just the product of two $SU(2)$ groups. 
Other interesting variation is given by models where instead of $SU(2)$ products one considers different subgroup dimensionalities, this is,  $\mG={ \prod_{\chi}} SU(N_\chi)_\chi$ with various $N_\chi$.  
Ref.~\cite{Fernandez:2015zsa} finds that, in models with very different dimensionalities $N_\chi$ --e.g., $\mG=SU(2)\times SU(100)$--, one naturally produces large hierarchies. 
We aim to discuss these ideas in a future work. 

\begin{acknowledgements}
We thank F. J. Llanes-Estrada and V.~Sanz for useful comments.
This work was supported by Grants FPU16/06960 (MECD), 
the MICINN (Spain) projects PID2019-107394GB-I00/AEI/10.13039/501100011033 (AEI/FEDER, UE) and PID2019-108655GB-I00/AEI/10.13039/501100011033 (AEI), 
the EU STRONG-2020 project [grant no. 824093], and STMS Grant from COST Action CA16108. JARC acknowledges support by Institut Pascal at Université Paris-Saclay during the Paris-Saclay Astroparticle Symposium 2021, with the support of the P2IO Laboratory of Excellence (program “Investissements d’avenir” ANR-11-IDEX-0003-01 Paris-Saclay and ANR-10-LABX-0038), the P2I axis of the Graduate School Physics of Université Paris-Saclay, as well as IJCLab, CEA, IPhT, APPEC, the IN2P3 master projet UCMN and EuCAPT.
This research was supported by the Munich Institute for Astro- and Particle Physics (MIAPP) which is funded by the Deutsche Forschungsgemeinschaft (DFG, German Research Foundation) under Germany´s Excellence Strategy – EXC-2094 – 390783311.
\end{acknowledgements}
 
\appendix 

\section{Impact of $\lambda_{LH}$ corrections on the $\eta$--$\varphi$ hierarchy} 
\label{app:lambdaLH-corrections}

In the case with $g_X=0$, we find that the effective potential has several critical points, being the global minimum given by one of them.
We restrict the analysis to the $\varphi,\, \eta \geq 0$ quadrant. Due to the $(\varphi\leftrightarrow-\varphi\, , \, \eta\leftrightarrow - \eta)$ symmetry of the Lagrangian, the remaining three quadrants are mirror copies of this one. 

Two of these critical points are placed at the $\eta$ and $\varphi$ axes,  
\begin{eqnarray}
(\varphi=0\, ,\, \eta)\, ,&& \qquad \mbox{with } \left(-\frac{99}{27}  +  \log   \left(\frac{g_H^2 \eta ^2 }{4 \mu ^2}\right)+\frac{128 \pi^2}{27} \frac{ \lambda_H}{g_H^4}\right) \,=\, 0\, ,
\\
(\varphi\, ,\, \eta=0)\, ,&& \qquad \mbox{with } \left(-\frac{99}{27}  +  \log   \left(\frac{g_L^2 \varphi ^2 }{4 \mu ^2}\right)+\frac{128 \pi^2}{27} \frac{ \lambda_L}{g_L^4}\right) \,=\, 0\, .
\end{eqnarray}

The remaining critical points depend on the value of $\lambda_{LH}$ and have $\varphi,\, \eta\neq 0$. They are given by the system of equations,   
\begin{eqnarray}
\lambda_{LH}&=&\, -\frac{27g_L^4 \varphi ^2}{64 \pi ^2 \eta ^2} \left(-\frac{99}{27} +    \log
   \left(\frac{g_L^2 \varphi ^2}{4 \mu ^2}\right)+\frac{128 \pi^2}{27}  \frac{\lambda_L}{g_L^4}\right) \, ,
\label{eq:rel1}\\ 
\lambda_{LH}&=&\,  -\frac{27g_H^4 \eta ^2}{64 \pi ^2 \varphi ^2} \left(-\frac{99}{27}  +  \log
   \left(\frac{g_H^2 \eta ^2 }{4 \mu ^2}\right)+\frac{128 \pi^2}{27} \frac{ \lambda_H}{g_H^4}\right)\, .
   \label{eq:rel2}
   \end{eqnarray}

In the case $\lambda_{LH}=0$ the two brackets on the r.h.s. of Eqs.~(\ref{eq:rel1}) and~(\ref{eq:rel2}) are zero and, therefore, equal. This identity provides the relation (in Eq.~(\ref{eq:hierarchy1})), 
\begin{equation}
\log\mathfrak{R} = \frac{128\pi^2}{27}\left(\frac{\lambda_L}{g_L^4}-\frac{\lambda_H}{g_H^4}\right)\ , 
\label{eq:logR0}
\end{equation}
with the hierarchy ratio $\mathfrak{R}\equiv  \frac{g_H^2\eta^2}{g_L^2\varphi^2}$.

This relation is modified in the case with $\lambda_{LH}\neq 0$, where the two brackets on the r.h.s. of Eqs.~(\ref{eq:rel1}) and~(\ref{eq:rel2}) are not zero any longer and are in general different. Nonetheless, 
dividing~(\ref{eq:rel2}) by~(\ref{eq:rel1}) is now allowed and leads to the relation, 
\begin{eqnarray}
1&=&\, \frac{g_H^4 \eta^4}{g_L^4\varphi^4}  
\frac{ \left(-\frac{99}{27}  +  \log
   \left(\frac{g_H^2 \eta ^2 }{4 \mu ^2}\right)+\frac{128 \pi^2}{27} \frac{ \lambda_H}{g_H^4}\right)  }{\left(-\frac{99}{27} +    \log
   \left(\frac{g_L^2 \varphi ^2}{4 \mu ^2}\right)+\frac{128 \pi^2}{27}  \frac{\lambda_L}{g_L^4}\right)}\, .
   \end{eqnarray}
It is not difficult to rewrite this expression in terms of $\mathfrak{R}$ and $\lambda_{LH}$ in the form,  
\begin{eqnarray}
\log{\mathfrak{R}} &=& \frac{128\pi^2}{27}\left(\frac{\lambda_L}{g_L^4}-\frac{\lambda_H}{g_H^4}\right) 
\,+\, 
\frac{64 \pi ^2}{27} \frac{\lambda_{LH}}{g_L^2g_H^2}
\left(\mathfrak{R}-\frac{1}{\mathfrak{R}}\right) \, .
   \label{eq:logR-modif}
\end{eqnarray}

In general, the system in Eqs.~(\ref{eq:rel1}) and~(\ref{eq:rel2}) has either one or three real critical point solutions $(\varphi>0\, ,\, \eta>0)$, which depend on the value of $\lambda_{LH}$.  
In order to understand the physical meaning of these solutions and how they evolve with $\lambda_{LH}$, we now discuss an example with fixed values $\{\lambda_L=0.001, g_L=0.6, \lambda_H=-0.001, g_H=0.6\}$. We chose coupling values that allow a clear enough visualisation of the potential minimum evolution. Other values show the same behaviour but transitions are much quicker and not as illustrative. These values generate a local minimum placed above the bisector $\varphi=\eta$. Through the interchanges $g_L\leftrightarrow g_H$ and $\lambda_L\leftrightarrow \lambda_H$, we would obtain the symmetric case, with a minimum under the quadrant bisector.

\begin{figure}[!t]
\begin{center}
\includegraphics[scale=0.65]{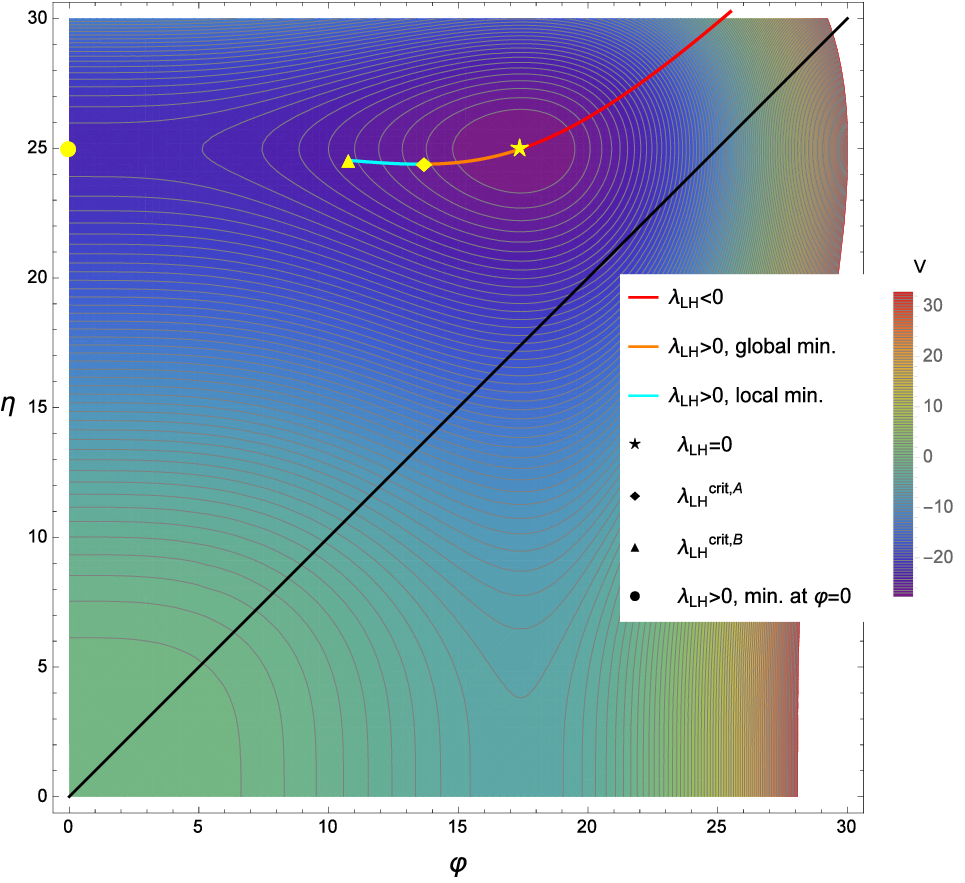}
\caption{{\small Effect of the $\lambda_{LH}$ parameter in the position of the minimum. In the background, a contourplot of the potential $V(\varphi,\eta)$ is represented for $\{\lambda_{LH}=0, \lambda_L=0.001, g_L=0.6, \lambda_H=-0.001, g_H=0.6\}$, with its minimum ($\star$). 
For $\lambda_{LH}<0$, the minimum of the potential is slowly displaced (red curve) towards the diagonal (black line). 
For $\lambda_{LH}>0$, apart from the original minimum, there is one additional local minimum on each axis $\varphi=0$ ($\bullet$) and $\eta=0$.
As one increases $\lambda_{LH}$, the central minimum remains at first the global one but its position is displaced towards the $\varphi=0$ axis (orange line). As the parameter increases, for $\lambda_{LH}>\lambda_{LH}^{crit,A}$ ($\blacklozenge$), the central minimum turns into a local minimum instead of the global one (blue line).  
When $\lambda_{LH} > \lambda_{LH}^{crit,B}$ ($\blacktriangle$), the central (local) minimum dissapears and only the minima at the axes remain. 
}}
\label{fig:contourplot}
\end{center}
\end{figure}

Fig.~\ref{fig:contourplot} shows the effect of including a $\lambda_{LH}\neq0$ parameter in the potential. First of all, in the background, we represent a contour-plot of the potential $V(\varphi,\eta)$ for the initial case $\lambda_{LH}=0$ with 
the previous choice of parameters.
The hierarchy is not very large in order to properly visualise the variations in the plot. 
We have an initial hierarchy $\mathfrak{R}>1$ with the minimum ($\star$) in a position $\langle\eta\rangle>\langle\varphi\rangle$. If we include a $\lambda_{LH}\neq0$ coupling this situation changes, but the effect of this parameter is quite different depending on  its sign. For $\lambda_{LH}<0$  (red line), the minimum of the potential is slowly displaced towards the diagonal bisector (black line), but the shape of the potential does not change with respect to the $\lambda_{LH}=0$ case. Therefore, the hierarchy decreases as $\lambda_{LH}$ gets more and more negative, approaching $\mathfrak{R}=1$ for $\lambda_{LH}\to \infty$. However, for $\lambda_{LH}>0$, both the position of the minimum and the shape of the potential change. 
In addition to the original central minimum, the two critical points at the $\varphi=0$ and $\eta=0$ axes also become local minima. 
Furthermore, one saddle point appears, respectively, next to each axis minima. Each of the two saddle points is placed between the corresponding axis minimum and the central minimum. Initially the central minimum is the global one, but the potential gets further and further distorted  as the value of $\lambda_{LH}$ increases, resulting on the three different regimes represented in Fig.~\ref{fig:contourplot}: 
\begin{itemize}

\item  
In the first one, the central minimum remains the global one but it is displaced towards the $\varphi=0$ axis (orange line), i.e. the hierarchy $\mathfrak{R}$ is increased but remains finite.  
The system stays in this regime until the parameter $\lambda_{LH}$ is increased up to a value $\lambda_{LH}^{crit,A}$ ($\blacklozenge$).

\item   
In the second one, the central minimum becomes less deep (while those in the axes are $\lambda_{LH}$ independent and remain fixed). It eventually turns simply into a local minimum instead of the global one, which is now located on the $\varphi=0$ axis ($\bullet$). The central minimum continues with the same displacement as before (blue line) so the associated hierarchy is still growing. In addition, during these two regimes, we have two more critical points, which are saddle points located near the minima at the axes,  
where the position of the saddle point close to the $\eta$ axis approaches the central minimum while increasing $\lambda_{LH}$. 
We stay in this regime for values from $\lambda_{LH}^{crit,A}$  ($\blacklozenge$) up to $\lambda_{LH}^{crit,B}$ ($\blacktriangle$).

\item   
When we reach this value $\lambda_{LH}^{crit,B}$ we enter the third regime: the saddle point reaches the central (local) minimum, resulting on the disappearance of both critical points. For higher values of $\lambda_{LH}$ only the minima at the axes remain; we could say that in this regime, the finite hierarchy that we previously had disappears, and we go to an {\it infinite hierarchy regime}, $\varphi=0\, , \,  \eta>0$ ($\bullet$).

\end{itemize}

\begin{figure}[!t]
\begin{center}
\includegraphics[scale=0.65]{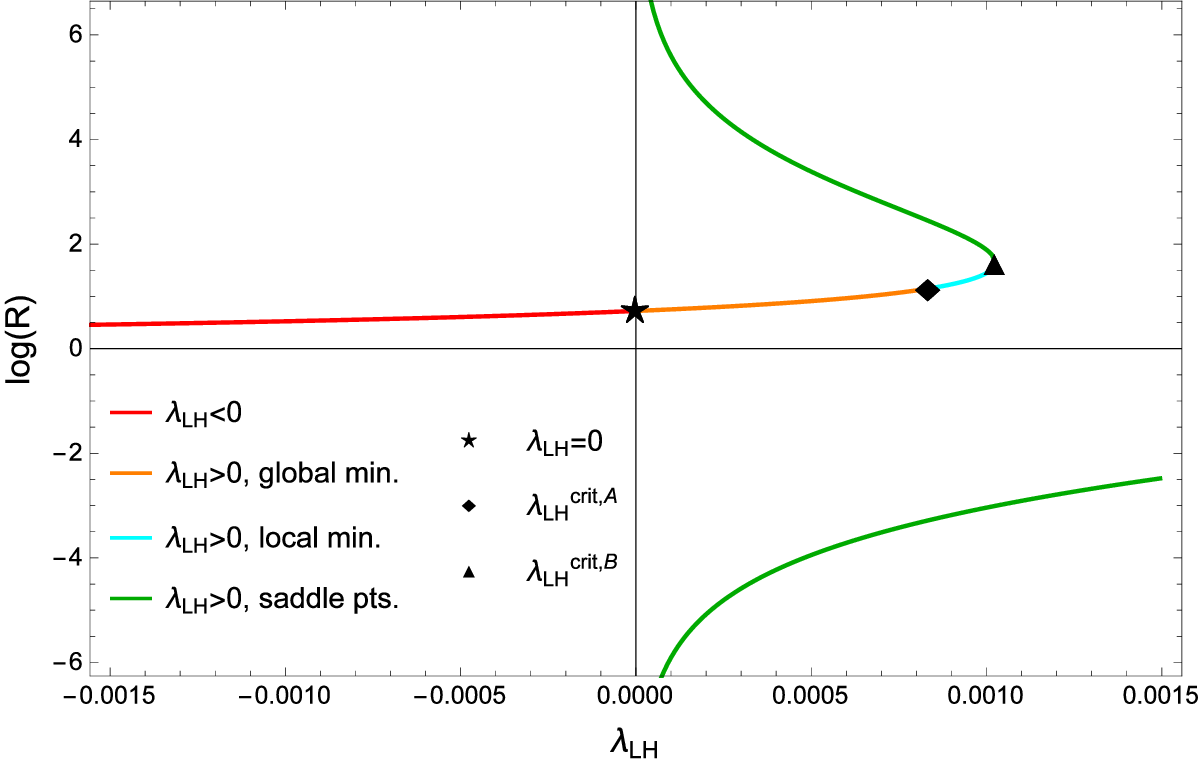}
\caption{{\small Hierarchy $\mathfrak{R}$ regimes depending on the value of $\lambda_{LH}$. For the initial case $\{\lambda_{LH}=0, \lambda_L=0.001, g_L=0.6, \lambda_H=-0.001, g_H=0.6\}$ we have a hierarchy depicted by $\star$. For $\lambda_{LH}<0$ the hierarchy decreases slowly, as represented by the red line. On the other hand, for $\lambda_{LH}>0$ it increases as shown by the orange line for the case of the central minimum being the global one up to $\lambda_{LH}^{crit,A}$ ($\blacklozenge$) and by the blue line when the central minimum is a local one. When parameter $\lambda_{LH} > \lambda_{LH}^{crit,B}$ ($\blacktriangle$), the central (local) minimum disappears and the system enters into the {\it infinite hierarchy regime}. 
The green lines represent the hierarchy associated to the saddle points (one next to each axis) that appear as additional solutions to Eq.~(\ref{eq:logR-modif}) 
but do not represent a physical hierarchy between the $L-H$ sectors.
}}
\label{fig:fases}
\end{center}
\end{figure}

The behaviour of the potential can also be visualised through Fig.~\ref{fig:fases}, where we show the value of $\mathfrak{R}$ for the solutions of Eq.~(\ref{eq:logR-modif}) as a function of $\lambda_{LH}$ for the critical points with $\varphi,\, \eta\neq 0$. 
Note that 
there are two additional critical points placed at the $\varphi=0$ and $\eta=0$ axes, not depicted in Fig.~\ref{fig:fases}. 
One can observe the same regimes described before but in terms of $\mathfrak{R}$. 
The colours and markers in this figure are the same as in Fig.~\ref{fig:contourplot}. We can see that for $\lambda_{LH}<0$, there is only one critical point and the hierarchy slowly decreases as $\lambda_{LH}$ gets more and more negative. On the other hand, for $\lambda_{LH}>0$, one finds at first three critical points: one local minimum and two saddle points (green lines). The hierarchy of the central local minimum increases up to a maximum value $\lambda_{LH}^{\rm crit,\, B}$ ($\blacktriangle$), which corresponds to the situation in which the central minimum merges with one of the saddle points and disappears. 
Beyond this point, there is only one critical point (a saddle point near the $\eta=0$ axis) and we move to the {\it infinite hierarchy regime}, with the global minimum $\varphi=0\, , \,  \eta>0$ ($\bullet$ in Fig.~\ref{fig:contourplot}). 
The green lines represent the saddle points (one next to each axis). These critical points appear as  ($\varphi,\eta$) solutions of the system of equations (\ref{eq:rel1}) and (\ref{eq:rel2}), additionally to the central minimum for $0<\lambda_{LH}<\lambda_{LH}^{\rm crit,\, B}$.

In summary, we find the scenario with $\lambda_{LH}\neq 0$ very compelling with a very rich phenomenology. The appearance of different phases with a metastable vacuum in some ranges or the presence of an infinite hierarchy deserve further investigation but the topic goes beyond the scope of this article.

\section{Probability $\mathfrak{P}^{(\alpha_L)}_{\rm cumul}$  from $\mO(g_X^2)$ corrections to the effective potential} 
\label{app:gX-corrections}

In this Appendix we provide the analytical expression of the cumulative probability (for hierarchies from $\mathfrak{R}_0$ up to $\infty$) when we fix the $\alpha_L=\{g_L^2,\lambda_L\}$ and $\{Q_L g_X, Q_H g_X\}$ parameters, with $g_X\neq 0$: 
\begin{eqnarray}
\mathfrak{P}^{(\alpha_L)}_{\rm cumul}  
&=&  
 \frac{\left(1+\mathfrak{a} \right)^3}{6\left(1+\mathfrak{b}\right)} 
 \left(\frac{27\ln\mathfrak{R}_0}{32\pi} -\frac{4\pi\lambda_L}{g_L^4}  + \mathfrak{c}\right)^{-2}
\, , 
\label{eq:cond-prob-cumul_gXnot0}
\end{eqnarray}
with the $\mathcal{O}(g_X^2)$ terms, 
\begin{eqnarray}
&&\qquad\qquad \mathfrak{a}=  \frac{Q_H^2 g_X^2}{g_L^2}\bigg[   
- \frac{9 g_L^2}{128\pi^2}\left(1+ 2 \ln\mathfrak{R}_0\right)
+\frac{2}{3}  - \frac{g_X^2}{64\pi^2}\left(19(Q_L^2-Q_H^2) + 6 Q_H^2 \ln\mathfrak{R}_0 \right) 
+ \frac{2\lambda_L}{3g_L^2}
\bigg]\,,
\nonumber\\
&&\qquad\qquad
\mathfrak{b}= \frac{ 2 Q_H^2g_x^2}{3g_L^2} \,, \qquad 
   { \mathfrak{c}=\frac{3g_X^2}{32\pi g_L^2}\left(19 Q_L^2-22Q_H^2 + 6Q_H^2\ln\mathfrak{R}_0 \right)  }\,,
\end{eqnarray}
for $\ln\mathfrak{R}_0\geq  \frac{128\pi^2}{27}\left(\frac{\lambda_L}{g_L^4}+ \frac{\epsilon_{CW}}{4\pi\epsilon_{g^2}}\right) \, (1+\mathcal{O}(g_X^2)) $.  
The form for smaller $\mathfrak{R}_0$ can also be derived in a straightforward way. It is easy to check that Eq.~(\ref{eq:cond-prob-cumul_gXnot0}) turns into~(\ref{eq:cond-prob-cumul_gX=0}) in the small $g_X$ limit. For this approximation to be valid one needs $g_X^2/g_L^2\ll 1$, $g_X^2/(4\pi)\ll 1$ and  $g_X^2 \ln\mathfrak{R}_0/(4\pi)\ll 1$. 
Hence, even for small $g_X$ ($g_X^2\ll 4\pi$ and $g_X^2\ll g_L^2$), we expect Eq.~(\ref{eq:cond-prob-cumul_gXnot0}) to fail for large enough $\ln\mathfrak{R}_0$, so its asymptotic expansion for $\mathfrak{R}_0\to\infty$ is not provided.

\bibliography{bibliography_CW}

\begin{thebibliography}{10}
\providecommand{\url}[1]{{#1}}
\providecommand{\urlprefix}{URL }
\expandafter\ifx\csname urlstyle\endcsname\relax
  \providecommand{\doi}[1]{DOI \discretionary{}{}{}#1}\else
  \providecommand{\doi}{DOI \discretionary{}{}{}\begingroup
  \urlstyle{rm}\Url}\fi

\bibitem{Coleman:1973jx}
S.R. Coleman, E.J. Weinberg, Phys. Rev. \textbf{D7}, 1888 (1973).
\newblock \doi{10.1103/PhysRevD.7.1888}

\bibitem{Casas:1994qy}
J.A. Casas, J.R. Espinosa, M.~Quiros, Phys. Lett. \textbf{B342}, 171 (1995).
\newblock \doi{10.1016/0370-2693(94)01404-Z}

\bibitem{Buttazzo:2013uya}
D.~Buttazzo, G.~Degrassi, P.P. Giardino, G.F. Giudice, F.~Sala, A.~Salvio,
  A.~Strumia, JHEP \textbf{12}, 089 (2013).
\newblock \doi{10.1007/JHEP12(2013)089}

\bibitem{Ellis:1986yg}
J.R. Ellis, K.~Enqvist, D.V. Nanopoulos, F.~Zwirner, Mod. Phys. Lett. A
  \textbf{1}, 57 (1986).
\newblock \doi{10.1142/S0217732386000105}

\bibitem{Barbieri:1987fn}
R.~Barbieri, G.F. Giudice, Nucl. Phys. B \textbf{306}, 63 (1988).
\newblock \doi{10.1016/0550-3213(88)90171-X}

\bibitem{Ciafaloni:1996zh}
P.~Ciafaloni, A.~Strumia, Nucl. Phys. B \textbf{494}, 41 (1997).
\newblock \doi{10.1016/S0550-3213(97)00138-7}

\bibitem{Casas:2014eca}
J.A. Casas, J.M. Moreno, S.~Robles, K.~Rolbiecki, B.~Zald\'\i{}var, JHEP
  \textbf{06}, 070 (2015).
\newblock \doi{10.1007/JHEP06(2015)070}

\bibitem{Weinberg:1988cp}
S.~Weinberg, Rev. Mod. Phys. \textbf{61}, 1 (1989).
\newblock \doi{10.1103/RevModPhys.61.1}

\bibitem{Bardeen:1995kv}
W.A. Bardeen, in \emph{{Ontake Summer Institute on Particle Physics}} (1995)

\bibitem{Fichet:2012sn}
S.~Fichet, Phys. Rev. D \textbf{86}, 125029 (2012).
\newblock \doi{10.1103/PhysRevD.86.125029}

\bibitem{Cabrera:2008tj}
M.E. Cabrera, J.A. Casas, R.~Ruiz~de Austri, JHEP \textbf{03}, 075 (2009).
\newblock \doi{10.1088/1126-6708/2009/03/075}

\bibitem{Ghilencea:2012qk}
D.M. Ghilencea, G.G. Ross, Nucl. Phys. B \textbf{868}, 65 (2013).
\newblock \doi{10.1016/j.nuclphysb.2012.11.007}

\bibitem{Weinberg:1973am}
E.J. Weinberg, {Radiative corrections as the origin of spontaneous symmetry
  breaking}.
\newblock Ph.D. thesis, Harvard U. (1973).
\newblock
  \urlprefix\url{http://inspirehep.net/record/85345/files/arXiv:hep-th_0507214.pdf}

\bibitem{Gildener:1976ih}
E.~Gildener, S.~Weinberg, Phys. Rev. \textbf{D13}, 3333 (1976).
\newblock \doi{10.1103/PhysRevD.13.3333}

\bibitem{Chataignier:2018kay}
L.~Chataignier, T.~Prokopec, M.G. Schmidt, B.~Świeżewska, JHEP \textbf{08},
  083 (2018).
\newblock \doi{10.1007/JHEP08(2018)083}

\bibitem{Khoze:2016zfi}
V.V. Khoze, A.D. Plascencia, JHEP \textbf{11}, 025 (2016).
\newblock \doi{10.1007/JHEP11(2016)025}

\bibitem{Fernandez:2015zsa}
G.~Garc\'\i{}a~Fern\'andez, J.~Guerrero~Rojas, F.J. Llanes-Estrada, Nucl. Phys.
  B \textbf{915}, 262 (2017).
\newblock \doi{10.1016/j.nuclphysb.2016.12.010}.
\newblock [Erratum: Nucl.Phys.B 949, 114755 (2019)]

\end{thebibliography}

\bibliographystyle{spphys}

\end{document}